\documentclass[twocolumn,english,aps,pra,showpacs]{revtex4-1}
\usepackage[T1]{fontenc}
\usepackage[latin1]{inputenc}
\usepackage{amsmath}
\usepackage{graphicx}
\usepackage{amssymb}
\usepackage{esint}

\makeatletter
\@ifundefined{textcolor}{}
{%
 \definecolor{BLACK}{gray}{0}
 \definecolor{WHITE}{gray}{1}
 \definecolor{RED}{rgb}{1,0,0}
 \definecolor{GREEN}{rgb}{0,1,0}
 \definecolor{BLUE}{rgb}{0,0,1}
 \definecolor{CYAN}{cmyk}{1,0,0,0}
 \definecolor{MAGENTA}{cmyk}{0,1,0,0}
 \definecolor{YELLOW}{cmyk}{0,0,1,0}
 }

\@ifundefined{definecolor}
 {\@ifundefined{definecolor}
 {\@ifundefined{definecolor}
 {\usepackage{color}}{}
}{}
}{}
\makeatother

\makeatother

\usepackage{babel}

\makeatother

\usepackage{babel}

\makeatother

\usepackage{babel}

\makeatother

\usepackage{babel}

\makeatother

\usepackage{babel}

\begin{document}

\title{Density distribution of a trapped two-dimensional strongly
interacting Fermi gas}

\author{Alexey A. Orel$^{1}$, Paul Dyke$^{1,2}$, Marion Delehaye$^{1,3}$,
Chris J. Vale$^{1}$, and Hui Hu$^{1}$}

\affiliation{$^{1}$Centre for Atom Optics and Ultrafast Spectroscopy, Swinburne University of Technology,
Melbourne 3122, Australia \\
 $^{2}$Department of Physics and Astronomy, and Rice Quantum Institute,
Rice University, Houston, TX 77251, USA \\
$^3$Departement de Physique, Ecole Normale Superieure, 24 rue
Lhomond, 75005 Paris, France}

\date{\today}
\begin{abstract}
We calculate and measure the density distribution and cloud size of
a trapped two-dimensional $^{6}$Li Fermi gas near a Feshbach resonance
at low temperatures. Density distributions and cloud sizes
are calculated for a wide range of interaction parameters using a local
density approximation (LDA) and a zero-temperature equation of state
obtained from quantum Monte Carlo simulations reported by G. Bertaina and S. Giorgini, Phys. Rev. Lett.
\textbf{106}, 110403 (2011). We find that LDA predictions
agree well with experimental measurements across a Feshbach resonance. Theoretical results for Tan's contact
parameter in a trapped gas are reported along with predictions for static structure factor
at large momentum which could be measured in future Bragg
spectroscopy experiments on two-dimensional Fermi gases. 
\end{abstract}


\maketitle

\section{Introduction}

The striking ability to manipulate and control ultracold atomic $^{6}$Li
and $^{40}$K Fermi gases has allowed the experimental investigation
of strongly interacting two-component Fermi gases \cite{rmpBDZ,rmpGPS}.
Using highly anisotropic pancake-shaped potentials to
confine atoms in the lowest axial mode \cite{Turlapov2D,Vale2D,Kohl2D},
it has become possible to realize experimentally two-dimensional (2D)
Fermi systems.  At low temperatures these can exhibit exotic properties such
as Berezinskii-Kosterlitz-Thouless (BKT) \cite{BKT,ZhangPRA2008,TemperePRA2009}
or inhomogeneous Fulde-Ferrell-Larkin-Ovchinikov (FFLO) superfluidity
\cite{FFLO,RiceFFLO,HLDFFLO,FFLO2D}. It has also been proposed that
a 2D interacting Fermi gas may provide useful insights into high-temperature
superconductivity \cite{MiyakePTP1983,RanderiaPRL1989} and itinerant
ferromagnetism \cite{mitFerroMagnetism,ConduitPRA2010}. To date,
a weakly interacting 2D Fermi gas has been imaged in situ \cite{Turlapov2D}
and used to characterize the crossover from two to three-dimensions \cite{Vale2D}.
The observation of 2D confinement induced resonances and measurement of the molecular binding energy using rf-spectroscopy of a strongly interacting Fermi gas was recently reported \cite{Kohl2D}.  However, the thermodynamic properties of strongly interacting Fermi gases and fermionic mixtures \cite{NishidaPRL2008} in reduced dimensions are yet to be fully explored.

On the theoretical side, numerous studies of 2D Fermi gases have been
presented, addressing superfluid transitions
\cite{ZhangPRA2008,PetrovPRA2003}, the effects of harmonic trapping \cite{TormaPRL2005},
and population and mass imbalance \cite{TemperePRA2009,TemperePRB2007,ConduitPRA2008}.
Of particular importance, the equation of state of a 2D uniform Fermi gas was recently
obtained through quantum Monte-Carlo (QMC) simulations \cite{Giorgini2D}
through the crossover from a Bardeen-Cooper-Schrieffer (BCS) superfluid of
Cooper pairs to a Bose-Einstein condensate (BEC) of tightly-bounded
molecules \cite{rmpGPS}. Tan's universal many-body contact parameter
was also found in 2D using the adiabatic relation \cite{Giorgini2D,TanRelations}.
While the most theoretical studies have relied on a perturbative or mean-field
approach, the ab-initio QMC results at zero temperature \cite{Giorgini2D}
should provide a quantitative description of the many-body ground state
of a strongly interacting 2D Fermi gas.

In this work, we measure the density distribution and cloud size of
a two-dimensional trapped $^{6}$Li Fermi gas at low temperatures
in the BEC-BCS crossover and compare the data with theoretical predictions.
The theoretical density distribution is calculated using a local density
approximation (LDA) \cite{LiuPRA2007}, based on the zero temperature
QMC equation of state \cite{Giorgini2D}. We find good qualitative
agreement between experiment and theory in the strongly interacting
regime near Feshbach resonance. When the atom number becomes large, we observe also substantial
deviation from the 2D equation of state when new transverse vibrational modes are populated.  

We also give theoretical predictions for Tan's contact parameter in a trapped gas
and the static structure factor at large momentum, which could be measured
in future Bragg spectroscopy experiments on 2D Fermi gases. Nontrivial pair correlations
are reflected in the many-body part of the contact parameter. The many-body contribution to the contact exhibits a maximum near the 3D Feshbach resonance. In the deep BEC limit, however, the two-body contribution to the contact arising from the molecular state dominates.

This paper is structured as follows: In the next section, we introduce
the LDA and the QMC results for the zero-temperature equation of state
for a uniform strongly interacting Fermi gas. In Sec. III, we discuss
the production of a strongly interacting trapped 2D Fermi gas of $^{6}$Li
atoms and how to calculate the density distribution and cloud size
from the QMC equation of state within LDA. The experimental procedure
for the density measurements is briefly summarized. In Sec. IV, we
present the theoretical density distributions and sizes and compare
these with the experimental measurements. In Sec. V. we find
Tan's contact and the large-momentum static structure factor of a 2D trapped
Fermi gas. Conclusions and future perspectives are given in Sec. VI.

\section{LDA and 2D uniform Fermi gas in the BEC-BCS crossover}

The equation of state of a strongly interacting Fermi gas in \emph{homogeneous}
space provides a convenient way to calculate the density distribution in a harmonic trap using the
local density approximation (LDA) \cite{LiuPRA2007}. The basic idea is that for a sufficiently large
number of particles in a slowly varying trapping potential $V_{ext}\left({\bf r}\right)$
we may treat the trapped Fermi system as a collection of many \emph{independent}
units that behave locally as a uniform Fermi gas. The correlation
between different units, for example, the surface energy of each unit,
is assumed to be negligibly small. Therefore, the local chemical potential
of a unit at position ${\bf r}$ may be written as, 
\begin{equation}
\mu\left({\bf r}\right)\equiv\mu\left[n\left({\bf r}\right)\right]=\mu-V_{ext}\left({\bf r}\right),
\end{equation}
where $n\left({\bf r}\right)$ is the local density and $\mu$ is
the chemical potential at the trap center. At zero temperature the local chemical potential
$\mu\left[n\left({\bf r}\right)\right]$ of the locally uniform unit
depends on the local density $n\left({\bf r}\right)$ only. Hence,
given the local equation of state $\mu[n({\bf r})]$ at position
${\bf r}$, we could solve inversely the density $n\left({\bf r}\right)=n(\mu-V_{ext}({\bf r)})$.
The chemical potential at the trap center $\mu$ is set by the normalization condition $\int d{\bf r}n\left({\bf r}\right)=N$, where $N$ is the total number of particles. The LDA has been shown
to work well in a wide range of situations \cite{LiuPRA2007}.
It is valid for either non-interacting or strongly interacting Fermi
gases in different geometries from 3D to 1D.

The essential ingredient of the LDA is the local uniform equation of state
$\mu(n)$. For a non-interacting two-component (spin-1/2) 2D Fermi
gas at zero temperature, the chemical potential is simply the Fermi
energy $\mu=E_{F}=\hbar^{2}k_{F}^{2}/(2m)$, where $m$ is the mass
of fermions and the 2D Fermi wave-vector is given by $k_{F}=(2\pi n)^{1/2}$.
Therefore, the non-interacting chemical potential is proportional
to the density, $\mu=\pi\hbar^{2}n/m$. The mean energy per particle is
$E/N=\hbar^{2}k_{F}^{2}/(4m)=E_{F}/2$.

For an interacting 2D Fermi gas in the BEC-BCS crossover, the energy
per particle $E/N$ has been calculated by Bertaina and Giorgini as a function of the interaction
strength, by using the fixed-node diffusion Monte Carlo method \cite{Giorgini2D}.
Here, we extract the chemical potential from the QMC data of the energy
per particle, since $\mu=\partial E/\partial N$. A suitable parameterization
of the QMC data for $E/N$ is therefore needed, as we discuss
in detail below.

In the 2D BEC-BCS crossover, a peculiar feature of the contact interactions
(between two fermions with unlike spins) is that any attraction, however
small, will support a two-particle bound state with energy $\epsilon_{B}=-4\hbar^{2}/[\exp\left(2\gamma\right)ma_{2D}^{2}]$, where $\gamma\simeq0.577216$ is the Euler's constant and $a_{2D}$
is the 2D $s$-wave scattering length \cite{RanderiaPRL1989,LiuPRB2010}.
This is in sharp contrast with the 3D BEC-BCS crossover, where a
two-body bound state appears on only one side of the Feshbach resonance where the 3D scattering length is positive \cite{rmpGPS}. The scattering length in 2D $a_{2D}$ is always {\em positive} due to the existence of the bound state.
The unitarity limit with an infinitely large scattering length ($a_{2D}\rightarrow+\infty$)
is in fact trivial: it corresponds simply to the non-interacting (BCS)
limit. In the opposite (BEC) limit with infinitely small scattering
length ($a_{2D}\rightarrow0^{+}$), where the energy of the bound
state is infinitely large, two fermions are tightly bound to form
a composite molecule. There will be a repulsive interaction between
two composite molecules, characterized by an effective scattering length
$a_{d}>0$.

The interaction strength in 2D may be expressed as a dimensionless
interaction parameter $\eta=\ln(k_{F}a_{2D})$. The weakly interacting
BEC and BCS limits correspond to $\eta\rightarrow-\infty$ and $\eta\rightarrow+\infty$,
respectively. The strongly interacting crossover regime occurs at
about $\eta=0$, where $a_{2D}\sim k_{F}^{-1}$.

We interpolate the QMC data for the 2D equation of state $(E/N-\epsilon_{B}/2)/E_{FG}$
in the BCS-BEC crossover with a smooth, continuous analytical function
$f(\eta)$ that consists of three parts. On the BEC side in the range
$\eta<-1/2$ we use the equation of state for a gas of composite molecules
with a molecular scattering length $a_{d}$ \cite{Giorgini2D}: \begin{eqnarray}
\frac{E}{N_{d}}-\epsilon_{B} & = & \frac{2\pi\hbar^{2}n_{d}}{m_{d}}\frac{1}{\ln\left[1/na_{d}^{2}\right]}\times\nonumber \\
 &  & \left\{ 1-\frac{\ln\ln\left[1/n_{d}a_{d}^{2}\right]-\left(\ln\pi+2\gamma+1/2\right)}{\ln\left[1/n_{d}a_{d}^{2}\right]}\right\} ,\label{energyBEC}\end{eqnarray}
 where $m_{d}=2m$ and $N_{d}=N/2$ are respectively the mass and
number of molecules, and $n_{d}=n/2$ is their density. By assuming
that $a_{d}/a_{2D}=\alpha_{m}\approx0.6$ \cite{Giorgini2D}, Eq.
(\ref{energyBEC}) turns into \begin{equation}
f_{BEC}\left(\eta<-\frac{1}{2}\right)\simeq\frac{0.5}{3.55-2\eta}\left[1-\frac{\ln\left(3.55-2\eta\right)-2.80}{3.55-2\eta}\right].\label{fyitaBEC}\end{equation}
 On the BCS side, for $\eta>2.72$ we consider a Padé-type approximate
function \begin{equation}
f_{BCS}\left(\eta>2.72\right)=\frac{1+a_{1}\eta^{-1}+a_{2}\eta^{-2}}{1+b_{1}\eta^{-1}+b_{2}\eta^{-2}},\label{fyitaBCS}\end{equation}
 where the Padé coefficients $a_{1}=0.164106,a_{2}=0.702385$ and
$b_{1}=1.16411,b_{2}=2.40527$ are obtained by minimizing the standard
deviation between the QMC data and the values of the function (\ref{fyitaBCS})
under the constraint of $f_{BCS}(\eta\rightarrow+\infty)\rightarrow1-1/\eta$,
which is required in a weakly interacting normal Fermi liquid in 2D.
In the crossover regime between the BEC and BCS limits we use a sixth order
polynomial function \begin{equation}
f_{crossover}\left(-\frac{1}{2}\leq\eta\leq2.72\right)=\sum_{i=0}^{6}c_{i}\eta^{i}.\label{fyitaCrossover}\end{equation}
 The coefficients are selected to provide the fit best to the the QMC data and to
ensure continuity of the equation of state function and of its first and second
derivatives at the two connection points $\eta=-0.5$ and $\eta=2.72$.
We find that, $c_{0}=0.200219$, $c_{1}=0.154862$, $c_{2}=-0.0144822$,
$c_{3}=0.070831$, $c_{4}=-0.01977$, $c_{5}=0.00891172$, and $c_{6}=-0.00108548$.

\begin{figure}[htp]

\begin{centering}
\includegraphics[clip,width=0.45\textwidth]{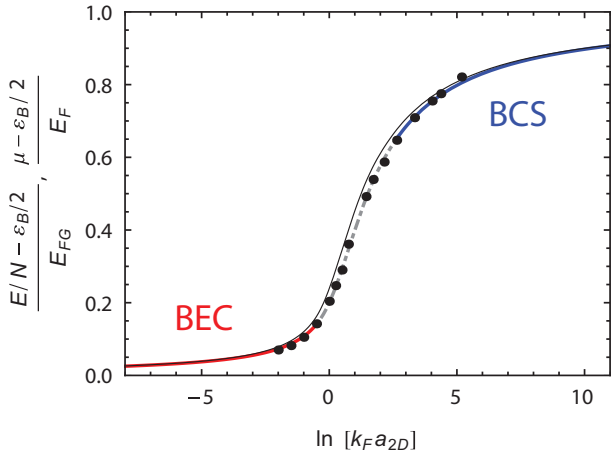} 
\par\end{centering}

\caption{(color online) Interpolating energy function $f(\eta)$ for the equation
of state of a 2D Fermi gas in the BEC-BCS crossover. The red solid
line on the BEC side is a fit corresponding to an equation of state
of a gas of composite bosons, Eq. (\ref{fyitaBEC}). The blue line
on the BCS side is the Padé-type approximation function of Eq. (\ref{fyitaBCS})
that minimizes the standard deviation with respect to the QMC data
\cite{Giorgini2D}. The dot-dashed line in the crossover regime is
a polynomial fit ensuring the continuity of the function and its first and second
derivatives at the two connection points, see Eq. (\ref{fyitaCrossover}).
The black circles show the QMC data of the equation of state. The
chemical potential derived from the interpolating energy function
is plotted by the thin black line.}

\label{fig1} 
\end{figure}

As shown in Fig. 1, the interpolating function $f(\eta)$ for the equation of state provides and excellent fit to the QMC data. By using $\mu=\partial E/\partial N$,
we find that, \begin{equation}
\frac{\mu-\epsilon_{B}/2}{E_{F}}\equiv f_{\mu}(\eta)=f(\eta)+\frac{1}{4}\frac{df(\eta)}{d\eta}.\label{mu}\end{equation}
 The dimensionless chemical potential $f_{\mu}(\eta)$ is shown in
Fig. 1 by the thin black line.

\section{A 2D strongly interacting Fermi gas in harmonic traps}

\subsection{Achieving the 2D regime}

For an atomic Fermi gas in a harmonic trap, the 2D regime is achieved
if the chemical potential $\mu$ and temperature $T$ are sufficiently
small compared to the excitation energy in one dimension ($z$).
We consider a spin-1/2 Fermi gas of $N$ $^{6}$Li atoms
with equal spin-populations in a highly oblate harmonic trap, \begin{equation}
V_{ext}^{3D}\left({\bf r}\right)=\frac{1}{2}m\left[\omega_{x}^{2}x^{2}+\omega_{y}^{2}y^{2}+\omega_{z}^{2}z^{2}\right],\end{equation}
 where $\omega_{x}\approx\omega_{y}=\omega_{\perp}$ and $\omega_{z}$
are the trapping frequencies in the radial ($x,y$) and axial ($z$)
directions, respectively. The trap aspect ratio $\lambda=\omega_{z}/\omega_{\perp}\gg1$.

The basic requirement for achieving a 2D trapped Fermi gas may be estimated
by considering first the zero-temperature and non-interacting limits,
in which the properties in a 2D harmonic trap, \begin{equation}
V_{ext}\left({\bf r}\right)=m\omega_{\perp}^{2}\left(x^{2}+y^{2}\right)/2=m\omega_{\perp}^{2}\rho^{2}/2,\end{equation}
can be conveniently understood using LDA. As $\mu=\pi\hbar^{2}n/m$
in an ideal 2D uniform Fermi gas, we obtain that $n({\bf r})=m\mu({\bf r})/(\pi\hbar^{2})=m[\mu-V_{ext}\left({\bf r}\right)]/(\pi\hbar^{2})$.
In other words, we expect a Thomas-Fermi (TF) distribution, \begin{equation}
n({\bf r})=n_{TF}\left(1-\frac{\rho^{2}}{\rho_{TF}^{2}}\right),\label{dstyIG}\end{equation}
 where $n_{TF}$ is the TF peak density and $\rho_{TF}$ is the TF
radius. Once the distance $\rho$ is larger than the TF radius, the
density is necessarily zero. The TF peak density and radius are related
to the chemical potential by $\mu=\pi\hbar^{2}n_{TF}/m$ and $\mu=m\omega_{\perp}^{2}\rho_{TF}^{2}/2$,
respectively. Using the normalization condition $\int d{\bf r}n\left({\bf r}\right)=N$,
it is straightforward to obtain that, $n_{TF}=N^{1/2}/(\pi a_{\perp}^{2})$
and $\rho_{TF}=(4N)^{1/4}a_{\perp}$, where $a_{\perp}\equiv[\hbar/(m\omega_{\perp})]^{1/2}$
is the characteristic oscillator length in the radial direction. The
2D chemical potential or Fermi energy $E_{F}$ is given by, $\mu=E_{F}=\sqrt{N}\hbar\omega_{\perp}$.
The characteristic Fermi temperature is $T_{F}=E_{F}/k_{B}$ and the
Fermi wave-vector $k_{F}$ in 2D harmonic traps is given by $k_{F}=[2mE_{F}/\hbar^{2}]^{1/2}=(4N)^{1/4}a_{\perp}^{-1}$.

As the lowest excitation energy in the $z$ direction is $\hbar\omega_{z}$,
one finds that the 2D regime can be reached if $\mu, E_F <\hbar\omega_{z}$
 and $T<\hbar\omega_{z}/k_{B}$. The former condition requires that the total number of atoms $N$ must be less than a 2D critical number, $N_{2D}$, equal to the number of single particle states with energy less than the lowest lying state with one transverse excitation.  It is straightforward to show that $N_{2D} = \lambda^2$.
In our experiment with $^{6}$Li atoms, the trapping frequencies are
$\omega_{z}\simeq2\pi\times2800$ Hz and $\omega_{\perp}\simeq2\pi\times47$
Hz, leading to $\lambda\approx60$, $a_{\perp}\approx6.0$ $\mu$m
and $N_{2D}\approx3600$.

\subsection{Achieving the strongly interacting regime}

Experimentally, the strongly interacting regime is reached by tuning
an external magnetic field $B$ near a Feshbach resonance ($B_{0}=834$
G) for $^{6}$Li atoms, for which the $s$-wave scattering length
\begin{equation}
a_{3D}\left(B\right)=a_{bg}\left(1+\frac{\Delta B}{B-B_{0}}\right)\left[1+\alpha\left(B-B_{0}\right)\right]\end{equation}
 can be changed precisely to arbitrary values \cite{Bartenstein05}. Here, $a_{bg}=-1405a_{0}$
with $a_{0}\approx0.529\times10^{-10}$ m is the background scattering
length, $\Delta B=300$ G is the width of Feshbach resonance and $\alpha=0.0004$
G$^{-1}$. In our highly oblate geometry, the tight-confinement in
the $z$-direction induces a bound state in the 2D $x-y$ plane. Therefore,
one can express the 2D scattering length in terms of the 3D scattering
length \cite{rmpBDZ,PetrovPRA2001},
\begin{equation}
a_{2D}=a_{z}\left(\frac{2\sqrt{\pi/b}}{e^{\gamma}}\right)\exp\left[-\sqrt{\frac{\pi}{2}}\frac{a_{z}}{a_{3D}}\right],\label{a2D}
\end{equation}
where $a_{z}\equiv[\hbar/(m\omega_{z})]^{1/2}$ and $b\approx0.915$.
In Fig. 2, we plot the dimensionless interaction parameter $\eta=\ln(k_{F}a_{2D})$
as a function of the magnetic field at the 2D critical number of atoms,
$N=N_{2D}=3600$. As indicated by the dotted line, at the location of the 3D Feshbach resonance
($B=B_{0}$), the interaction parameter $\eta\approx1$.

\begin{figure}[htp]

\begin{centering}
\includegraphics[clip,width=0.45\textwidth]{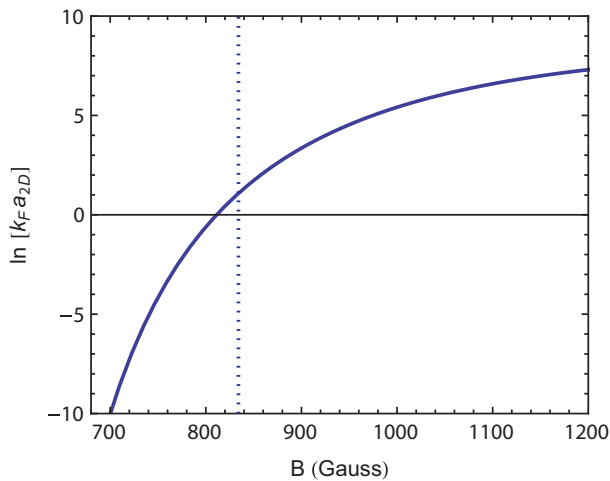} 
\par\end{centering}

\caption{(color online) The dimensionless interaction parameter $\ln(k_{F}a_{2D})$
of a trapped 2D interacting Fermi gas near a Feshbach resonance. The
vertical dotted line indicates the resonance position. Here, we calculate
the Fermi wave-vector $k_{F}$ at $N=N_{2D}=3600$ by using $k_{F}=(4N)^{1/4}a_{\perp}^{-1}$.
The 2D scattering length is calculated using Eq. (\ref{a2D}) with
$a_{z}\approx770$ nm for $\omega_{z}=2\pi\times2800$ Hz.}

\label{fig2} 
\end{figure}

\subsection{Theoretical density distributions}

Let us now consider the density distribution of an interacting 2D
Fermi gas in the BEC-BCS crossover within LDA. The simple relation
$n({\bf r})=m\mu({\bf r})/(\pi\hbar^{2})$, useful for the ideal gas,
is no longer applicable. We have to obtain numerically the local density
from the local chemical potential by using the $f_{\mu}$-function,
defined in Eq. (\ref{mu}). That is, we need to solve the following equation to find the
density $n({\bf r})$ from, \begin{eqnarray}
\mu\left({\bf r}\right) & = & \mu-m\omega_{\perp}^{2}\left(x^{2}+y^{2}\right)/2,\\
 & = & \frac{\epsilon_{B}}{2}+\frac{\pi\hbar^{2}}{m}n\left({\bf r}\right)f_{\mu}\left[\ln\sqrt{2\pi n\left({\bf r}\right)}a_{2D}\right].\end{eqnarray}
 In the inversion procedure, our analytic interpolating $f_{\mu}$-function
appears to be very convenient. The density distribution $n({\bf r})$
is calculated for an initially chosen value of chemical potential
$\mu$. We then adjust $\mu$ to satisfy the number equation $\int d{\bf r}n\left({\bf r}\right)=N$.
At the final stage, we quantify the cloud size using root mean square
(rms) radius, \begin{equation}
\sqrt{\left\langle \rho^{2}\right\rangle }=\left[\frac{\int d{\bf r}n\left({\bf r}\right)\left(x^{2}+y^{2}\right)}{\int d{\bf r}n\left({\bf r}\right)}\right]^{1/2}.\end{equation}

\begin{figure}[htp]

\begin{centering}
\includegraphics[clip,width=0.45\textwidth]{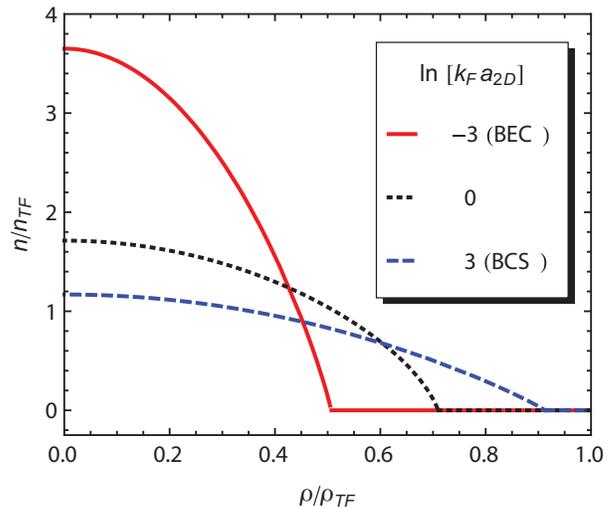} 
\par\end{centering}

\caption{(color online) Atomic density distributions of a 2D Fermi cloud at
three values of interaction strengths $\eta=\ln(k_{F}a_{2D})$. The
red solid line and blue dashed line correspond to the weakly interacting
BEC ($\eta=-3$) and weakly interacting BCS ($\eta=+3$) side. The
black dotted line stands for a strongly interacting crossover Fermi
gas ($\eta=0$).}

\label{fig3} 
\end{figure}

In Fig. 3, we plot the theoretical density distributions at three
values of the interaction parameter $\eta$, which correspond to the
weakly interacting BEC and BCS sides, and the strongly interacting
crossover regime. The density and distance from the center of
the cloud are plotted in units of the TF density $n_{TF}$ and the
TF radius $\rho_{TF}$, respectively. One finds that with decreasing
the interaction parameter $\eta$ from the BCS side to the BEC side,
the 2D cloud becomes denser and narrower in size,
as anticipated. On the BEC side, our analytic interpolating $f_{\mu}$-function
leads to the following asymptotic behavior, \begin{equation}
\frac{n\left(\rho\right)}{n_{TF}}=\sqrt{2\ln\left(\frac{2\pi}{\alpha_{m}}\right)-4\eta}-\left[2\ln\left(\frac{2\pi}{\alpha_{m}}\right)-4\eta\right]\frac{\rho^{2}}{\rho_{TF}^{2}},\end{equation}
 where $\alpha_{m}=a_{d}/a_{2D}\approx0.6$ is the ratio between 2D
molecular and atomic scattering length. Hence, with decreasing $\eta$
($\rightarrow-\infty$) the peak density increases as $[-4\eta]^{1/2}$
and the radius of the cloud decreases as $[-4\eta]^{1/4}$. On the
other hand, on the BCS side the density distribution converges to
the ideal Fermi gas result of Eq. (\ref{dstyIG}), \begin{equation}
\frac{n\left(\rho\right)}{n_{TF}}=\left(1+\frac{1}{2\eta}\right)-\left(1+\frac{1}{2\eta}\right)^{2}\frac{\rho^{2}}{\rho_{TF}^{2}}.\end{equation}
 In accord with these asymptotic density distributions, the cloud
sizes are given by, \begin{equation}
\frac{\left(\sqrt{\left\langle \rho^{2}\right\rangle }\right)_{BEC}}{\sqrt{\left\langle \rho^{2}\right\rangle _{IG}}}=\frac{1}{\left[2\ln\left(2\pi/\alpha_{m}\right)-4\eta\right]^{1/4}}\end{equation}
 and \begin{equation}
\frac{\left(\sqrt{\left\langle \rho^{2}\right\rangle }\right)_{BCS}}{\sqrt{\left\langle \rho^{2}\right\rangle _{IG}}}=\sqrt{\frac{2\eta}{\left(2\eta+1\right)}},\end{equation}
 in the BEC and BCS limits, respectively. Here, $\sqrt{\left\langle \rho^{2}\right\rangle _{IG}}=\rho_{TF}/\sqrt{3}$
is the rms cloud size of an ideal 2D Fermi gas. We have checked that
these analytic results agree well with our numerical calculations in the appropriate limits.  We have also checked the sensitivity of the calculated density profiles to the form of the equation of state.  The widths of the theoretical density profiles do not vary by more than 5$\%$ which is small on the scale of the width changes as the interaction strength is varied over the range considered here.

\subsection{Experimental measurements}

To measure the density distribution of a strongly interacting 2D Fermi
gas, we use an experimental setup similar to the one used in our previous
work \cite{Vale2D}. In brief, we start with a cloud of approximately
$1\times10^{5}$ $^{6}$Li atoms in the two hyperfine states $\left|F=1/2,m_{F}=\pm1/2\right\rangle $
in a far detuned 3D optical dipole trap. The cloud is evaporatively
cooled to the lowest possible temperature near the Feshbach resonance.
At this stage, the number of atoms is controlled by further lowering the trap depth
to spill atoms out of the dipole trap. We then ramp on a 2D optical trap in 200 ms
to create a highly oblate trap with trapping frequencies $\omega_{z}\simeq2\pi\times2800$
Hz and $\omega_{\perp}\simeq2\pi\times47$ Hz, which gives an aspect
ratio of approximately 60. Finally, we tune the interaction strength
by adiabatically ramping the external magnetic field to $810$ G,
$834$ G and $992$ G, where the cloud is held and imaged. The dimensionless
interaction parameters $\ln(k_{F}a_{2D})$ at these fields are about
$-0.5$, $+0.6$, and $+5$, respectively.

The critical number of atoms for reaching 2D regime is $N_{2D}\approx3600$.
Depending on the depth of the dipole trap, the final number of atoms
in the cloud can be varied in the range of $500$ to $5000$ atoms.
The final temperature of these small 2D and quasi-2D clouds is difficult to determine when interactions are present. However,
we anticipate it to be approximately $0.1\,T_{F}$ based on applying the same preparation procedure to clouds with a larger atom number.

Before the imaging, we allow a short time of flight ($500$ $\mu$s).
This allows us to resolve the density distribution in the tightly-confined
$z$-direction, since this time scale is long compared to $1/\omega_{z}\approx57$
$\mu$s. It is however much short compared to $1/\omega_{\perp}\approx3.4$
ms, and therefore, the cloud distribution in the radial direction
is essentially equivalent to the {\it in situ} profile. The imaging beam propagates roughly along
the radial $x$-direction, which means that there is an automatic
integration over the $x$-direction for the total density distribution.
We then integrate these distributions over the $z$-direction to generate a double-integrated
column density $\tilde{n}(y)$. The rms cloud size is calculated
from the first moment of the one-dimensional profile $\left\langle \rho^{2}\right\rangle = \int dy\tilde{n}(y{\bf )}y^{2}/\int dy\tilde{n}(y{\bf )}$.
Theoretically, we perform the same integration procedure in the $x$-direction
for the 2D density distribution. This should lead to the same distribution
as the experimentally measured double-integrated column density $\tilde{n}(y)$.

\section{Comparison between theory and experiment}

\begin{figure}[htp]

\begin{centering}
\includegraphics[clip,width=0.45\textwidth]{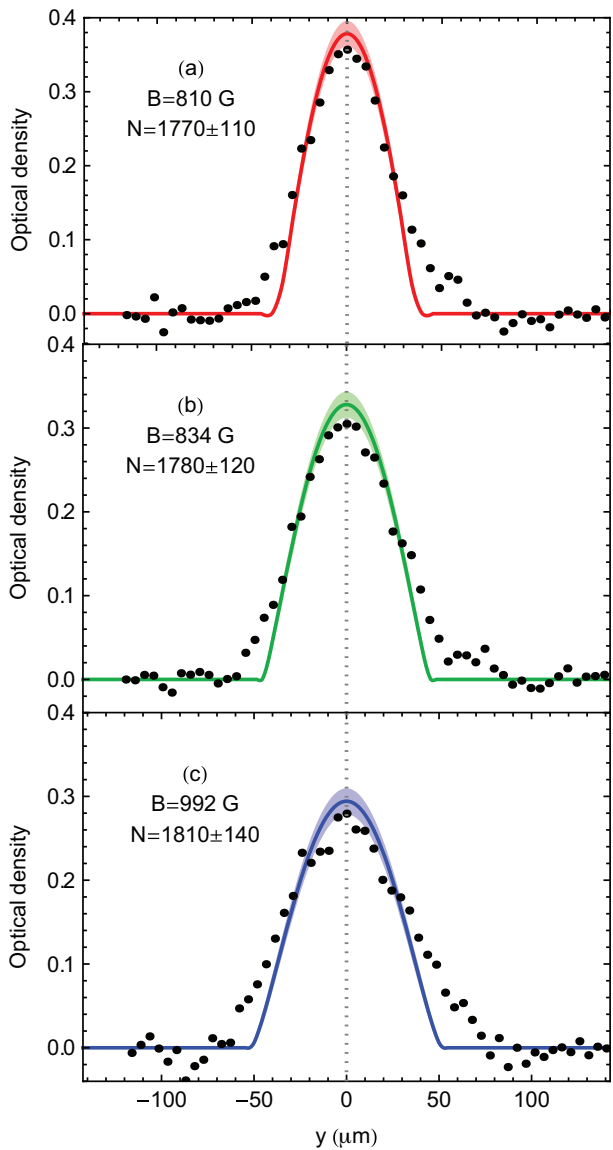} 
\par\end{centering}

\caption{(color online) Comparison between the theoretical LDA predictions
(lines) and experiment measurements (solid circles) for the column
density $\tilde{n}(y)$ at the magnetic fields, $810$ G (a), $834$
G (b), and $992$ G (c). The optical density is shown in arbitrary
unit as a function of the actual $y$-coordinate (in units of micrometer).}

\label{fig4} 
\end{figure}

In Fig. 4, we compare the LDA column density $\tilde{n}(y)$ (lines)
with the experimental measurements (solid circles) at three magnetic
fields. To reduce the experimental noise, we average the experimental
density distributions over many images with a range of atom numbers that are all well
below $N_{2D}$. Accordingly, the theoretical lines
correspond to the average number of atoms, while the standard deviation in
the atom number is illustrated by shaded region.  We observe good qualitative agreement between theory and experiment with no adjustable parameters.  The distributions become wider with increasing $\ln(k_F a_{2D})$ however, there are notable discrepancies between theory and experiment, particularly in the wings of the clouds.  This is due to a number of effects including the finite imaging resolution and recoil induced blurring during the imaging pulse.  The combination of these two artifacts is to lower the effective resolution of the imaging system to approximately 6 $\mu$m.  This alone however, is not enough to fully account for the observed discrepancies.  The remaining differences are most likely due to the finite temperature of the clouds which will show up most in the wings of the distribution.

\begin{figure}[htp]

\begin{centering}
\includegraphics[clip,width=0.45\textwidth]{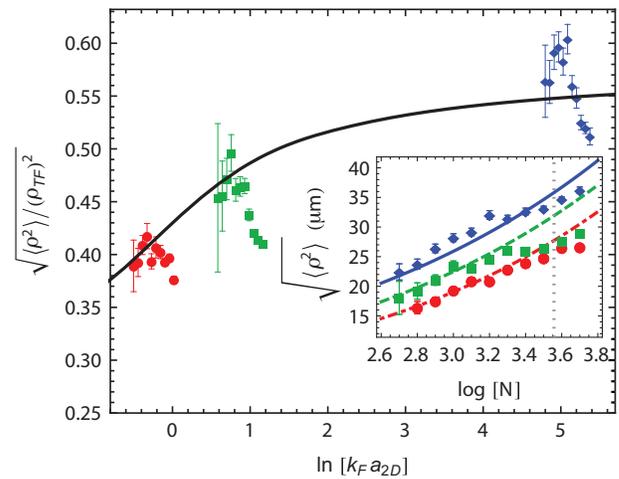} 
\par\end{centering}

\caption{(color online) The rms cloud size in units of the TF radius as a function
of the dimensionless interaction strength. The black line shows the
theoretical LDA prediction. The red circles, green squares, and blue
diamonds represent the experimental data at the magnetic fields, $810$
G, $834$ G, and $992$ G, respectively. The number of atoms at a
given magnetic field can be varied, giving rise to slightly different
interaction parameter. In the inset, the actual rms cloud sizes are
shown as a function of the decimal logarithm of the atom number. The blue
solid line (the highest) is the LDA prediction for 992 G; the green
dashed line (medium) is for 834 G; and the red dot-dashed line (the
lowest) is for 810 G. The vertical dotted line in
the inset indicates the position of the critical atom number.}

\label{fig5} 
\end{figure}

In Fig. 5, we plot the rms cloud size of the 2D Fermi gas in units
of the TF radius as a function of the interaction parameter $\eta=\ln k_{F}a_{2D}$.
The lines and symbols show, the LDA predictions and experimental
data, respectively. In the strongly interacting regime at the magnetic fields $B=810$
G and $834$ G, the LDA predictions agree quantitatively well with
the experimental data (solid red circles and green squares). At the high field $B=992$ G, the gas is more weakly interacting and the experimental
data (blue diamonds) lie slightly above the theory curve.  In the inset of Fig. 5, we present
the actual cloud size as a function of the number of atoms. The
agreement between theory and experiment near the Feshbach resonance
($810$ G and $834$ G) becomes more apparent.

In all three sets of measurements, the radial cloud size drops below the true 2D prediction for the largest atom numbers.  This happens when the first transverse excited state becomes energetically accessible and leads to a drop in the growth rate of the radial cloud size.  In this quasi-2D regime, shell structure associated with resolving the discrete transverse states can dramatically affect the density profiles \cite{Vale2D}.  The atom numbers for which the radial width departs from the 2D theory are slightly below what we would expect for an ideal Fermi gas.  This could be due to interactions or scattering resonances \cite{olshanii,PetrovPRA2001,haller, sala, peng} but also may arise through thermal excitations.

\section{Contact parameter and the static structure factor}

\subsection{Tan's contact parameter}

For a strongly interacting Fermi gas with contact interactions, the
asymptotic behavior of various physical quantities in the limit of
short-distance or large-momentum is governed by a single parameter,
called the contact, which measures the density of fermionic pairs
within a short distance. This was first discussed by Tan in 2008, when he derived
a set of exact universal relations for strongly interacting Fermi gases \cite{TanRelations}. Being an
important many-body parameter, Tan's contact
is also related to the thermodynamics via the adiabatic relation
\cite{TanRelations}.   The contact in a homogeneous 2D Fermi gas was calculated by Bertaina and Giorgini \cite{Giorgini2D}.  Here we use these results to find the contact and static structure factor in a trapped system.  In 2D, the adiabatic relation takes the form \cite{WernerPreprint},
\begin{equation}
{\cal I}=\frac{2\pi m}{\hbar^{2}}\frac{dE}{d\ln a_{2D}},\label{eq:Contact}\end{equation}
 where the derivative is taken at constant entropy. The calculation
of Tan's contact parameter for a 3D strongly interacting Fermi gas
was performed recently \cite{UniversalContact}, by using the similar
adiabatic relation.

The zero-temperature contact of a homogeneous 2D Fermi gas can be
calculated by substituting our interpolated energy per particle $E/N=\epsilon_{B}/2+E_{FG}f(\ln[k_{F}a_{2D}])$
into Eq. (\ref{eq:Contact}). We find that ${\cal I}={\cal I}_{2b}+{\cal I}_{mb}$,
where \begin{equation}
\frac{{\cal I}_{2b}}{Nk_{F}^{2}}=\frac{2\pi m}{\hbar^{2}k_{F}^{2}}\frac{d\left[\epsilon_{B}/2\right]}{d\ln a_{2D}}=\frac{8\pi}{e^{2\gamma}}\frac{1}{\left[k_{F}a_{2D}\right]^{2}}\end{equation}
 is the contribution from the two-body bound state and \begin{equation}
\frac{{\cal I}_{mb}}{Nk_{F}^{2}}=\frac{2\pi m}{\hbar^{2}k_{F}^{2}}\frac{d[E_{FG}f\left(\eta\right)]}{d\ln a_{2D}}=\frac{\pi}{2}\frac{df}{d\eta}\end{equation}
 is the contribution from the many-body correlations, respectively.
The two-body contact ${\cal I}_{2b}$ increases monotonically from
the BCS to the BEC limit. In contrast, the many-body contact ${\cal I}_{mb}$
should exhibit a maximum at the crossover regime, according to the
behavior of the energy $f$-function (see Fig. 1). As shown in Fig.
6 by a thin line, the maximum of the many-body contact occurs at $\eta\sim0.8$,
which is roughly the position of the Feshbach resonance (see also,
ref. \cite{Giorgini2D}).

\begin{figure}[htp]

\begin{centering}
\includegraphics[clip,width=0.45\textwidth]{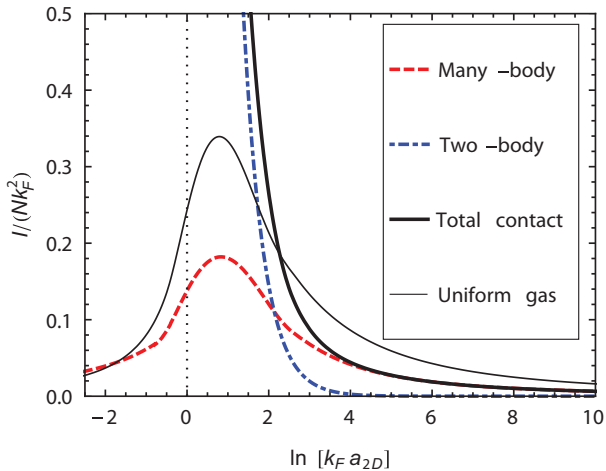} 
\par\end{centering}

\caption{(color online) Theoretical contact parameter of a trapped 2D Fermi
gas in the BEC-BCS crossover. The contact is shown in units of $Nk_{F}^{2}$.
The blue dot-dashed line and red dashed line show respectively the
contribution from the two-body bound state and the many-body part
(see text), while the thick black line gives the total contribution.
For comparison, we show by a thin line for the many-body part of the
contact of a homogeneous 2D Fermi gas.}

\label{fig6} 
\end{figure}

For a trapped interacting Fermi gas, we may calculate the contact
by using the LDA density distribution, \begin{equation}
{\cal I}_{T}=\int d{\bf r}\frac{{\cal I}({\bf r})}{\Delta V}=2\pi\int d{\bf r}\left[\frac{{\cal I}}{Nk_{F}^{2}}\right]\left({\bf r}\right)n^{2}\left({\bf r}\right),\end{equation}
 where we have summed the local contact density, ${\cal I}({\bf r})/\Delta V=[{\cal I}/(Nk_{F}^{2})]\left({\bf r}\right)\times n({\bf r})k_{F}^{2}({\bf r})$,
over the whole space. It is easy to see that, the two-body contact
is not affected by the density average, so that ${\cal I}_{T,2b}={\cal I}_{2b}$.
However, the many-body part may be significantly affected. In Fig.
6, we present the result for the contact of a trapped 2D Fermi gas
at the BEC-BCS crossover, with the many-body contact shown by a red
dashed line. It has roughly the same shape as the many-body contact
of a homogeneous gas, with a peak appearing at $\eta\sim0.8$. However,
the peak is about half as high due to the average over
the density distribution.

\subsection{Spin-antiparallel static structure factor}

Tan's contact for a 3D strongly interacting Fermi gas has been measured
in a number of ways. One appealing method is to measure the spin-antiparallel
static structure factor by using Bragg spectroscopy at large momentum
\cite{SwinTanWorks}, which has a $1/q$ tail with a prefactor given
by Tan's contact \cite{eplHLD}. In 2D, we may make a similar prediction.
It has been shown that the 2D pair correlation function $n^{(2)}\left({\bf r}\right)\propto{\cal I}\ln^{2}(\rho/a_{2D})$
\cite{WernerPreprint}. The $\ln^{2}(\rho/a_{2D})$ dependence can
be qualitatively understood from the two-body relative wave-function
$\psi_{rel}\left({\bf r}\right)\sim\ln(r/a_{2D})$, since $n^{(2)}\left({\bf r}\right)\propto\left|\psi_{rel}\left({\bf r}\right)\right|^{2}$.
The spin-antiparallel static structure factor is simply the Fourier
transform of the pair correlation function. Thus, we find that, 
\begin{eqnarray}
S_{\uparrow\downarrow}\left(q\gg k_{F}\right) & = & \frac{{\cal I}}{N}\frac{\left[\gamma+\ln\left(qa_{2D}\right)\right]}{\pi^{2}q^{2}},\\
 & = & \left[\frac{{\cal I}}{Nk_{F}^{2}}\right]\frac{\left[\gamma+\ln\left(k_{F}a_{2D}\right)+\ln\tilde{q}\right]}{\pi^{2}\tilde{q}^{2}},
 \end{eqnarray}
where $\tilde{q}\equiv q/k_{F}$ and $\gamma = 0.577216$. Compared with the 3D case, the
spin-antiparallel static structure factor in 2D decays faster with
increasing momentum $q$ ($q^{-2}$ compared to $q^{-1}$).

\begin{figure}[htp]

\begin{centering}
\includegraphics[clip,width=0.45\textwidth]{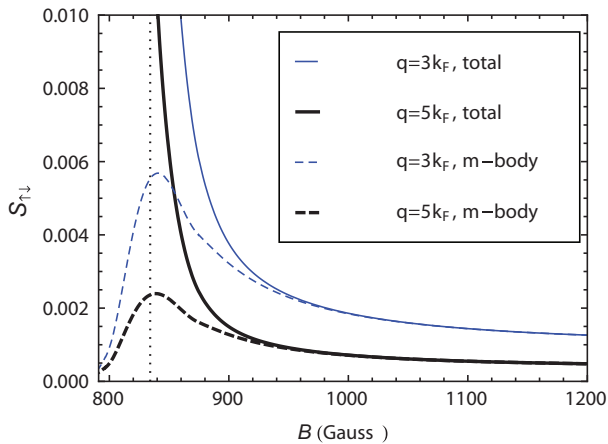} 
\par\end{centering}

\caption{(color online) The LDA prediction for the spin-antiparallel static
structure factor of a trapped 2D Fermi gas in the BEC-BCS crossover,
under our experimental conditions of $\omega_{z}\simeq2\pi\times2800$
Hz and $\omega_{\perp}\simeq2\pi\times47$ Hz for $N=N_{2D}=3600$
$^{6}$Li atoms. We show the many-body contribution by the dashed
lines. The thick and thin lines give respectively the predictions
at the transferred momentum $q=5k_{F}$ and $3k_{F}$.}

\label{fig7} 
\end{figure}

In Fig. 7 we plot the theoretical prediction for the spin-antiparallel static structure
factor of a trapped 2D Fermi gas in the BEC-BCS crossover for momenta $q=3k_{F}$ and
$q=5k_{F}$. We split the structure factor into the two-body and many-body
parts, in accord with the previous classification for the contact.
Near the Feshbach resonance, the static structure factor at $q=3k_{F}$
is about $0.02$, whose magnitude is accessible within current experimental
resolution \cite{SwinTanWorks}.

\section{Conclusions}

To conclude, we have predicted theoretically and measured experimentally
the density distribution and cloud size of a low-temperature two-dimensional harmonically trapped
Fermi gas in the BEC-BCS crossover. The theoretical
calculations have been carried out within a local density approximation,
based on the ab-initio zero-temperature equation of state obtained
from the fixed-node diffusion Monte Carlo simulations \cite{Giorgini2D}. The experimental
measurements were performed using a single two-dimensional Fermi
cloud of $^{6}$Li atoms near a Feshbach resonance. We have found
good qualitative agreement between theory and experiment.

We have also calculated an important many-body parameter, Tan's
contact, and have proposed that it can be straightforwardly measured
using Bragg spectroscopy for the spin-antiparallel static structure
factor as in three dimensions. In future studies, it will be interesting
to study both experimentally and theoretically the density distributions
at finite temperatures, which may elucidate the fermionic
Berezinskii-Kosterlitz-Thouless transition in two dimensions.

\begin{acknowledgments}
This work is supported by the Australian Research Council (ARC) Centre
of Excellence for Quantum-Atom Optics and by ARC Discovery Project
No. DP0984522. 
\end{acknowledgments}

\end{document}